\documentclass[10pt]{article}
\usepackage{mathrsfs}
\usepackage{amsmath,amsfonts,amssymb,mathtools}
\usepackage{cite}
\usepackage{dsfont}
\usepackage{graphicx}
\usepackage{hyperref}
\usepackage{verbatim}
\usepackage{slashed}
\usepackage[all]{xy}
\usepackage{xcolor}
\usepackage{subfig}
\usepackage{tikz}
\usetikzlibrary{shapes.geometric,arrows}
\usepackage{relsize}
\usepackage{caption}
\usepackage{float}
\usetikzlibrary{positioning}
\usepackage{geometry}
\usepackage[T1]{fontenc}
\usepackage{lmodern}
\usepackage{stackengine} 
 
\usepackage[utf8]{inputenc}

\hypersetup{
  colorlinks,
  citecolor=violet,
  linkcolor=blue,
  urlcolor=blue}

\csname @addtoreset\endcsname{equation}{section}
\DeclareMathAlphabet\mathbfcal{OMS}{cmsy}{b}{n}

\newcommand{\beq}{\begin{equation}}
\newcommand{\eeq}{\end{equation}}
\newcommand{\bea}{\begin{eqnarray}}
\newcommand{\eea}{\end{eqnarray}}
\newcommand{\ba}{\begin{array}}
\newcommand{\ea}{\end{array}}
\newcommand{\bit}{\begin{itemize}}
\newcommand{\eit}{\end{itemize}}
\newcommand{\nn}{\nonumber}

\newcommand{\complesso}{{\ \hbox{{\rm I}\kern-.6em\hbox{\bf C}}}}
\newcommand{\reale}{{\hbox{{\rm I}\kern-.2em\hbox{\rm R}}}}
\newcommand{\uno}{ \,  \raisebox{+0.14em}{{\hbox{{\rm \scriptsize ]}} \raisebox{-0.2em}{\kern-.8em\hbox{1}}}} \, }  

\newcommand{\p}{\partial}

\newcommand{\w}{\wedge}

\renewcommand{\a}{\alpha}

\newcommand{\g}{\gamma}

\renewcommand{\d}{\delta}
\newcommand{\D}{\Delta}

\newcommand{\Er}{{\mathbfcal{E}}}

\renewcommand{\k}{\kappa}

\renewcommand{\L}{\Lambda}

\newcommand{\m}{\mu}

\newcommand{\n}{\nu}
\renewcommand{\r}{\rho}
\newcommand{\s}{\sigma}

\renewcommand{\S}{\Sigma}

\newcommand{\om}{\omega}
\newcommand{\Om}{\Omega}

\begin{document}


\begin{titlepage}

\vspace{0.3cm}

\begin{flushright}
$LIFT$--11-4.25
\end{flushright}

\vspace{1.0cm}

\begin{center}
\renewcommand{\thefootnote}{\fnsymbol{footnote}}
\vskip 9mm  
{\Huge \bf Black holes in the external
\vskip 6mm
Bertotti-Robinson-Bonnor-Melvin
\vskip 10mm
   electromagnetic field}
\vskip 37mm
{\large {Marco Astorino$^{a}$\footnote{marco.astorino@gmail.com}
}}\\

\renewcommand{\thefootnote}{\arabic{footnote}}
\setcounter{footnote}{0}
\vskip 8mm
\vspace{0.2 cm}
{\small \textit{$^{a}$Laboratorio Italiano di Fisica Teorica (LIFT),  \\
Via Archimede 20, I-20129 Milano, Italy}\\
} \vspace{0.2 cm}
%

\end{center}

\vspace{2.0 cm}

\begin{center}
{\bf Abstract}
\end{center}
{An exact and analytical solution, in four-dimensional general relativity coupled with Maxwell electromagnetism, is built by means of a Lie point symmetry of the Ernst equations, the Harrison transformation. \\
The new spacetime describes a Schwarzschild-like black hole embedded into a general external back-reacting electromagnetic field, which is the superposition of the Levi-Civita-Bertotti-Robinson and the Bonnor-Melvin ones. The relation between the two homogeneous electromagnetic fields is clarified. Conserved charges and the first law of thermodynamics are analysed. Swirling generalisations are also considered. Limits to the known metrics such as Schwarzschild-Bertotti-Robinson, Schwarzschild-Bonnor-Melvin and Bertotti-Robinson-Bonnor-Melvin are discussed.}

\end{titlepage}

\addtocounter{page}{1}

\newpage


\section{Introduction}
\label{sec:introduction}

A number of consolidated experimental evidences confirm that astrophysical massive black holes at the center of galaxies are surrounded by huge (electro)magnetic fields \cite{Eatough:2013nva}. Therefore a detailed theoretical and analytical description that can model these objects would be very welcome. At the moment we know two families of exact solutions describing black holes in a back-reacting external (electro)magnetic field \cite{ernst-magnetic} and \cite{garcia-alekseev}, \cite{kerr-bertotti}. These represent black holes immersed into the Bonnor-Melvin universe \cite{bonnor}, \cite{melvin} or alternatively into the Levi-Civita-Bertotti-Robinson\footnote{Even if Levi-Civita was the first to discover this spacetime, in \cite{levi-civita}, several decades before Bertotti and Robinson, often he is omitted in the nomenclature; henceforward for brevity (and homogeneity with the standard notation) we might drop Levi-Civita's name too.\\ 
Moreover to avoid ambiguities we remark that Schwarzschild in the Levi-Civita-Bertotti-Robinson background has nothing to do with the work of \cite{maccallum}, recently named Schwarzschild in the Levi-Civita background. In fact this other Levi-Civita background is a cylindrical vacuum (non-electrovaccum) solution, also discovered by Levi-Civita but not in \cite{levi-civita}. \\
Note that those recently called, black holes in the Levi-Civita background are subcases of the black hole in the swirling background \cite{swirling} for the swirling parameter $\jmath=\infty$. This is obvious because they can be generated with the inversion transformation, which is a specialization of the Ehlers transformation, for large values of the Lie-point parameter (or the swirling parameter in this specific case), as shown in \cite{PD-NUTs}.} background  \cite{levi-civita}, \cite{bertotti}, \cite{robinson}. \\
In this article we want to extend, at the same time, both classes of these known solutions, in particular generalising their electromagnetic field. We would like to get a more general back-reacting external electromagnetic field, depending on several physical parameters, which can be dampened to recover the pure vacuum Schwarzschild solution. To do so, we plan to exploit a solution generating technique based on the symmetries of the Ernst equations for the Einstein-Maxwell theory. In particular we will use the Harrison transformation to add the Bonnor-Melvin background to the Schwarzschild black hole immersed into the Levi-Civita-Bertotti-Robinson background to get a Schwarzschild-like black hole in the electromagnetic field that is a superposition of the Bonnor-Melvin and the Levi-Civita-Bertotti-Robinson ones, as done in section \ref{sec:generation}. The generating methods are fundamental, in this context, because of the non-linearity of Einstein's field equations: these spacetimes would be very difficult to build, by solving directly the field equations. \\
Limits to already known subcases will be explicitly clarified; while limits to the novel electromagnetic backgrounds will be  also discussed, see section \ref{sec:Bertotti-Melvin}. These metrics, because they are continuously connected to the Schwarzschild metric, can be consider a good phenomenological description, or for a certain parametric range of the integrating constants, related to the external electromagnetic fields, or possibly, as a local model, that is, not far from the black hole.  \\
Further stationary generalisations can be obtained, in section \ref{sec:BH-in-Bertotti-Melvin-Swirling}, thanks to the Ehlers transformation, to add also a swirling background to the solution.\\

\section{Generation of the new solution: embedding Schwarzschild into Bertotti-Robinson-Bonnor-Melvin electromagnetic background}
\label{sec:generation}

\begin{figure}[h!]
\captionsetup[subfigure]{labelformat=empty}
\centering
\hspace{-1cm}
\subfloat[\hspace{0.1cm} $m=1$, $B=0.1$, $b=0.1$]{{\includegraphics[scale=0.5]{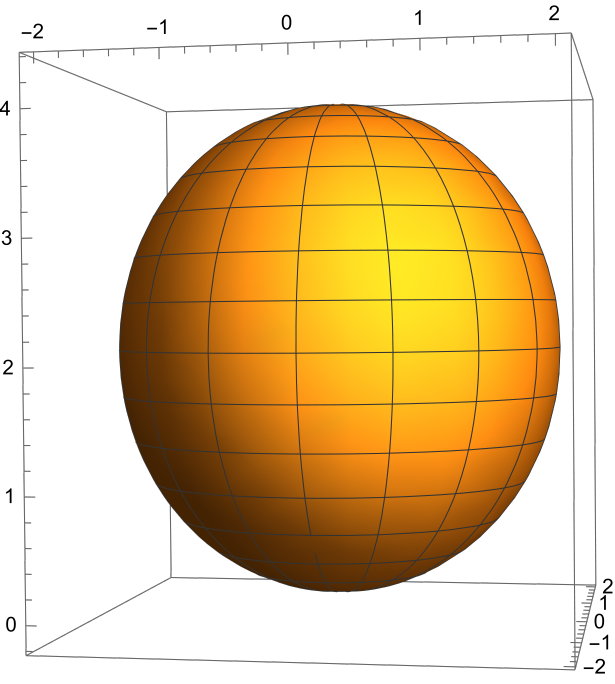}}}%
\subfloat[\hspace{0.8cm} $m=1$, $B=0.4$, $b=0.1$]{{ \hspace{0.5cm} \includegraphics[scale=0.4]{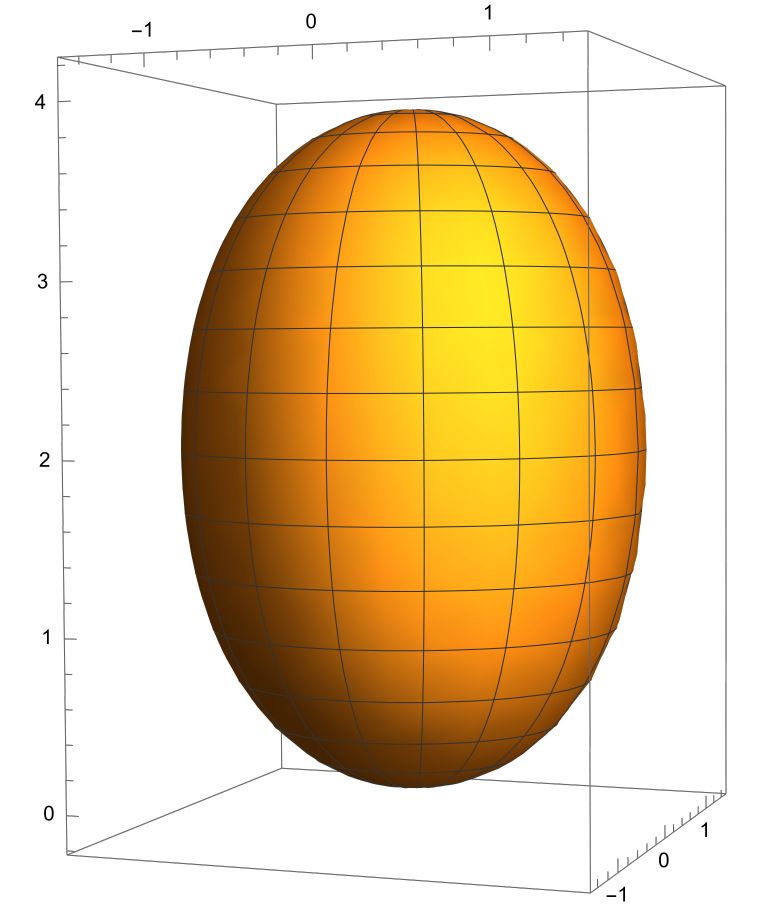}}}%
\subfloat[\hspace{0.8cm} $m=1$, $B=2$, $b=0.1$]{{\hspace{1cm}
\includegraphics[scale=0.5]{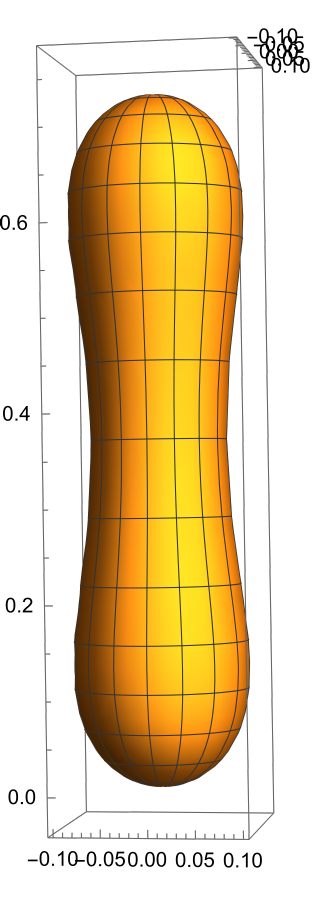}}}\\
\subfloat[\hspace{0.1cm} $m=1$, $B=0$, $b=0$]{{\includegraphics[scale=0.4]{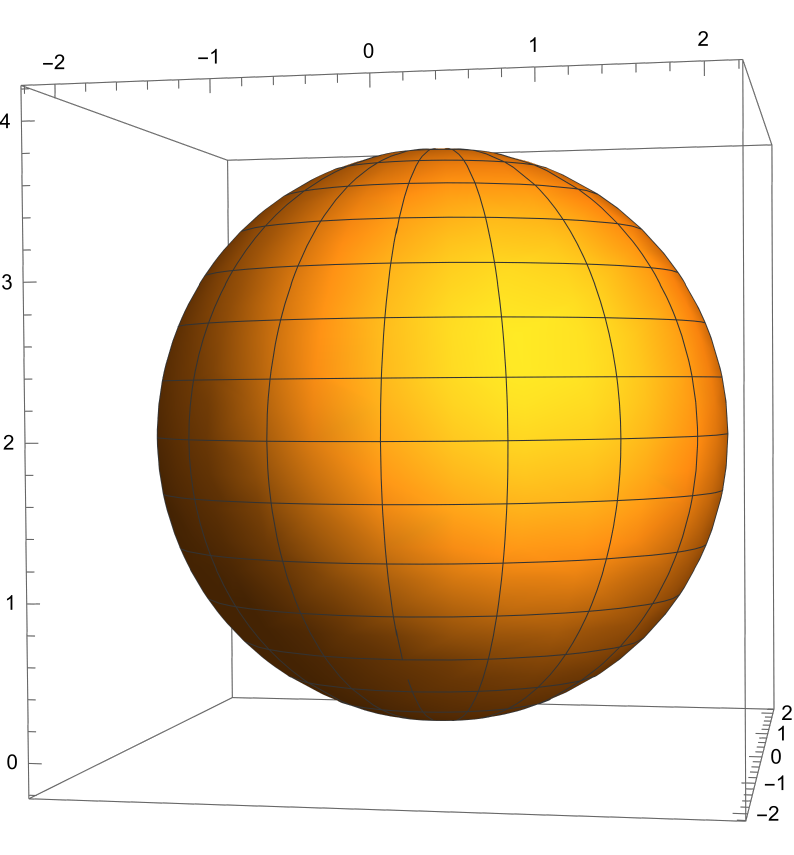}}}%
\subfloat[\hspace{0.8cm} $m=1$, $B=0.3$, $b=-0.4$]{{ \hspace{0.5cm} \includegraphics[scale=0.4]{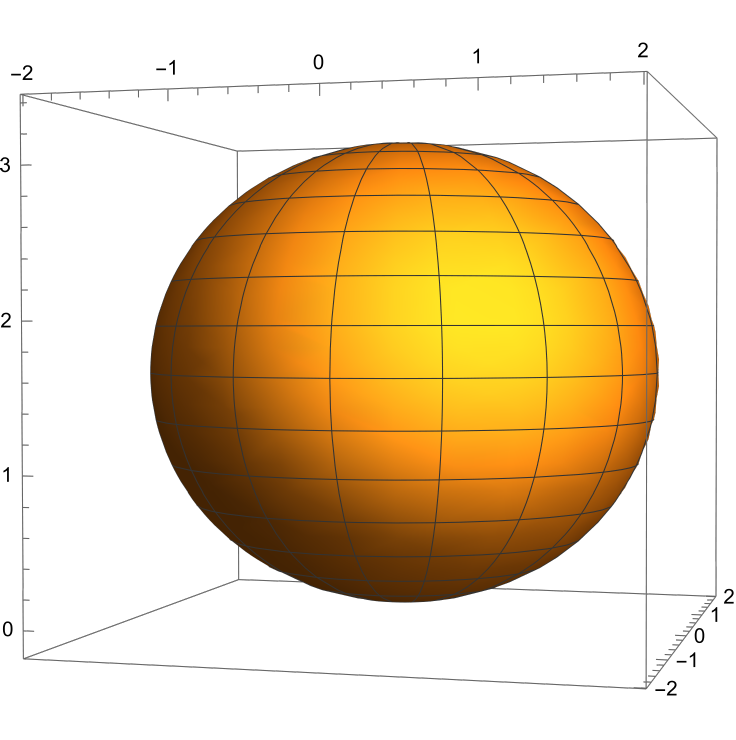}}}%
\subfloat[\hspace{0.8cm} $m=1$, $B=0.5$, $b=-0.6$]{{\hspace{-0.1cm}
\includegraphics[scale=0.4]{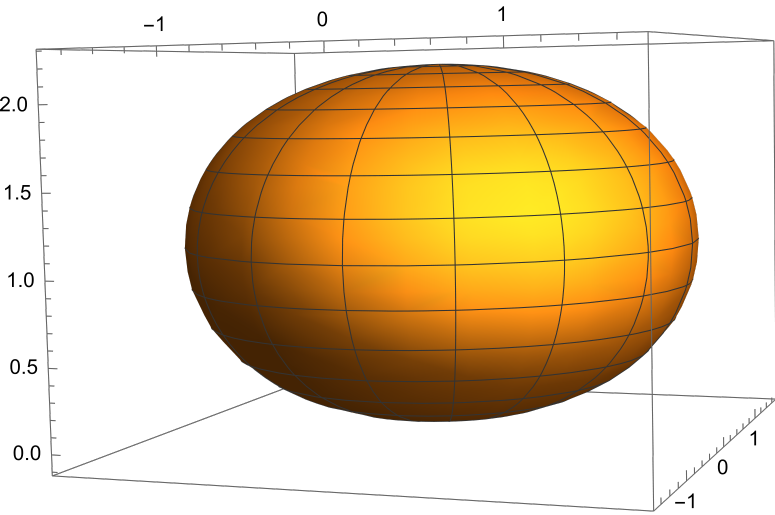}}}
\caption{\small Embedding in Euclidean three-dimensional space $\mathbb{E}^3$ of the event horizon of the black hole distorted by the presence of both the external Bertotti-Robinson and Bonnor-Melvin magnetic backgrounds, for different values of the parameters $b$ and $B$, while $m=1$. When $b=0$ and $B=0$ we have the spherical Schwarzschild horizon. More pictures can be found in figure \ref{fig:picture-horizons-more}.}%
\label{fig:picture-horizons}
\end{figure}

The theory we consider in this article is the standard Einstein general relativity in four-dimensions coupled with Maxwell electromagnetism. It is determined by the action
\beq
         I[g_{\m\n},A_\m] = - \frac{1}{16\pi G} \int_\mathcal{M} d^4x \sqrt{-g} \left(R - F_{\m\n}F^{\m\n} \right) \ . \nn
\eeq
By extremising the above action, with respect to the metric and the Maxwell potential, $g_{\m\n}$ and $A_\m$ respectively, one gets the Einstein-Maxwell field equations 
\bea  \label{field-eq-g}
                        &&   R_{\m\n} -   \frac{R}{2}  g_{\m\n} =   2 \left( F_{\m\r}F_\n^{\ \r} - \frac{1}{4} g_{\m\n} F_{\r\s} F^{\r\s}  \right) \quad ,   \\
       \label{field-eq-A}                  &&   \partial_\m ( \sqrt{-g} F^{\m\n}) = 0  \ \quad ,
\eea
where the Faraday field-strength is defined as $F_{\m\n}:= \nabla_\m A_\n - \nabla_\n A_\m $. \\
When one restricts to electromagnetic axisymmetric and stationary spacetimes, described generally by the Lewis-Weyl-Papapetrou (LWP) metric
\beq \label{LWP-magnetic}
                     ds^2 = -f ( d\varphi - \omega dt)^2 + f^{-1} \bigl[ \rho^2 dt^2 - e^{2\gamma}  \bigl( {d \rho}^2 + {d z}^2 \bigr) \bigr]      
\eeq
and by a vector gauge potential, compatible with the spacetime symmetries, $A_{\mu}=\left\{A_t(\r,z),0,0,A_\varphi(\r,z)\right\}$, Ernst had found in \cite{ernst2}, that the field equations  (\ref{field-eq-g})-(\ref{field-eq-A}) can be written as
\bea 
     \label{ee-ernst-ch}  \left( \textsf{Re} \ \Er + | \mathbf{\Phi} |^2 \right) \nabla^2 \Er   &=&   \left( \overrightarrow{\nabla} \Er + 2 \ \mathbf{\Phi^*} \overrightarrow{\nabla} \mathbf{\Phi} \right) \cdot \overrightarrow{\nabla} \Er   \quad ,       \\
     \label{em-ernst}   \left( \textsf{Re} \ \Er + | \mathbf{\Phi} |^2 \right) \nabla^2 \mathbf{\Phi}  &=& \left( \overrightarrow{\nabla} \Er + 2 \ \mathbf{\Phi^*} \overrightarrow{\nabla} \mathbf{\Phi} \right) \cdot \overrightarrow{\nabla} \mathbf{\Phi} \quad ;
\eea
where the complex electromagnetic and gravitational Ernst potentials, for the metric (\ref{LWP-magnetic}), are respectively defined by
\beq \label{def-Phi-Er} 
       \mathbf{\Phi} := A_\varphi + i \tilde{A}_t  \qquad , \qquad \qquad     \Er := f - \mathbf{\Phi} \mathbf{\Phi}^* + i h  \quad ,
\eeq
while $\tilde{A}_t$ and $h$ stem from
\bea
    \label{A-tilde-e} \overrightarrow{\nabla} \tilde{A}_t &:=&  f \r^{-1} \overrightarrow{e}_\varphi \times (\overrightarrow{\nabla} A_t + \omega  \overrightarrow{\nabla} A_\varphi ) \ \ , \\
    \label{h-e}    \overrightarrow{\nabla} h &:=& - f^2 \r^{-1} \overrightarrow{e}_\varphi \times \overrightarrow{\nabla} \omega - 2 \ \textsf{Im} (\mathbf{\Phi}^*\overrightarrow{\nabla} \mathbf{\Phi} )  \ \ .
\eea
The Ernst equations (\ref{ee-ernst-ch})-(\ref{em-ernst}) enjoy a set of eight (real) Lie-point symmetries, for more details see \cite{stephani-big-book} or \cite{enhanced}. Since we are interested in generalising the external electromagnetic field of the static uncharged black hole embedded into the Bertotti-Robinson background, we will consider the so-called Harrison transformation
\beq \label{harrison}
\Er \longrightarrow \bar{\Er} = \frac{\Er}{1-2\a^*\mathbf{\Phi} -\a\a^* \Er} \qquad \quad ,  \quad  \qquad  \mathbf{\Phi} \longrightarrow  \bar{\mathbf{\Phi}} = \frac{\mathbf{\Phi} + \a \Er}{1-2\a^*\mathbf{\Phi} -\a\a^* \Er} \quad .
\eeq
When this map is applied to a solution cast in the LWP form (\ref{LWP-magnetic})\footnote{On the other hand, if the Harrison transformation is applied to a seed expressed in the form of the conjugate of the (\ref{LWP-magnetic}), i.e. $ds^2 = -f ( dt - \omega d\varphi)^2 + f^{-1} \bigl[ \rho^2 d\varphi^2 + e^{2\gamma}  \bigl( {d \rho}^2 + {d z}^2 \bigr) \bigr]$, we would add electromagnetic monopoles, for details see \cite{Type-I}.}, it embeds the seed metric into the Bonnor-Melvin electromagnetic universe \cite{bonnor}, \cite{melvin}. As seed solution we chose the Schwarzschild black hole in the Bertotti-Robinson electromagnetic field, as the non-rotational limit\footnote{Note that the solution in \cite{garcia-alekseev} also represents a Schwarzschild black hole into the Levi-Civita-Bertotti-Robinson background, but written in Weyl coordinates which are a little more obscure than the one used in \cite{kerr-bertotti}. About their possible equivalence we refer to \cite{ortaggio}. It is worth clarifying that this class of type-D black holes, which we are considering as seed, is not contained in the Plebanski-Demianski family because the null directions of the Faraday tensor are not aligned with the two principal null directions of the Weyl tensor. The most general type-D black hole solutions, for the Einstein-Maxwell theory, where the null directions of the Faraday tensor is aligned with the null principal directions of the Weyl tensor, is discussed in \cite{most-D} and \cite{revisiting-D}.} of \cite{kerr-bertotti}, see also \cite{carminati}. The metric expressed in spherical coordinates takes the form
\beq \label{schwarschil-br-spherical}
      ds^2 = \frac{-\left( 1 -\frac{2m}{r} - m^2B^2 \right) \left( 1+ B^2r^2\right) dt^2 + \frac{dr^2}{\left(1 - \frac{2m}{r} - m^2B^2 \right) \left( 1+ B^2r^2\right)} +\frac{r^2 d\theta^2}{\D_\theta} + r^2 \sin^2\theta \D_\theta \D_\varphi^2 \, d\varphi^2}{1+B^2r^2-B^2r \cos^2 \theta (r-2m-rB^2m^2)}  
\eeq
with
\beq
         \D_\theta(\theta) = 1+m^2B^2\cos^2\theta \ .
\eeq
The constant $\D_\varphi$ is just a dilatation of the coordinate $\varphi$ and could be useful to properly define the period of the azimuthal angle to have a smooth geometry, at least outside the event horizon. The value that removes the conical singularity is given in eq. (\ref{Delta-phi}).\\
However, to apply the solution generating technique, it is more practical to cast the above metric into the LWP form in spherical-like coordinates $(t,r,x = \cos \theta, \varphi)$, as follows: 
\beq \label{LWP-spherical}
       ds^2 = -f ( \D_\varphi d\varphi - \omega dt)^2 + f^{-1} \left[ \rho^2 dt^2 - e^{2\gamma}  \left( \frac{dr^2}{\Delta_r} + \frac{d x^2}{\Delta_x} \right) \right] 
\eeq
where
\bea \label{f-seed}
        f(r,x) &=&  -\frac{r^2\Delta_x}{\Omega^2} \ , \hspace{3.2cm} \D_r(r) =  \, r(r-2m-rB^2m^2)(1+B^2r^2) \ , \\
        \gamma(r,x) &=& \frac{1}{2} \log \left( \frac{r^4 \D_x}{\Omega^4} \right) \ , \hspace{2cm} \D_x(x) = \,  (1-x^2)(1+B^2m^2x^2) \ , \nn \\
        \rho(r,x)   &=& \frac{\sqrt{\D_r \D_x}}{\Omega^2} \ ,\hspace{2.9cm} \Omega(r,x) = \, \sqrt{1+B^2r^2-B^2rx^2(r-2m-rB^2m^2)} \ , \nn   \\
         \omega(r,x) &=& 0 \ ,\label{w-seed}
\eea
while the magnetic field is given by the potential
\beq \label{A-seed-magn}
         A_\m = \left[ 0, 0, 0, \frac{\D_\varphi}{B} \left(\Omega - r \p_r\Omega -1 \right)  \right] \ .
\eeq
Note that the limit of the gauge potential for vanishing values of the external magnetic field, that is for $B \to 0$, is well defined and finite, even though it could be not apparent at first glance. \\
At this point we want to keep the system as simple as possible, so we are considering only magnetic fields. In this way we remain with a diagonal spacetime both for the initial seed solution and also for the transformed one. This second requirement is implemented by restricting to the real part only of the complex Harrison transformation, i.e. $\a=b/2$. Further generalisations also including electric fields and more general transformation with respect to the one in (\ref{harrison}) are considered in section \ref{sec:BH-in-Bertotti-Melvin-Swirling}. Note that in that case the mutual interaction of the electric seed of the solution and the magnetic field brought in by the Lie-point transformation makes the resulting black hole stationary rotating (even though the seed was static). This is a manifestation of the Lorentz force.   \\
From the seed solution and the definitions (\ref{def-Phi-Er})-(\ref{h-e}) we are able to derive the seed's $h$ and $\tilde{A}_t$, which are constants that we can set to zero without losing physical generality, then we have all the elements to write the seed Ernst complex potentials
\bea \label{Er-Phi}
       \mathbf{\Phi}  &=& \frac{1+mB^2rx^2-\Omega}{B\Om}  \ , \\
        \Er &=&  \frac{2}{B}\  \mathbf{\Phi}  \ .
\eea
Then the Harrison transformation (\ref{harrison}) adds to the seed solution an extra external Bonnor-Melvin magnetic field. The new solution is completely identified from the transformed Ernst potentials ($\bar{\Er},\bar{\Phi}$), as in (\ref{harrison}). In case we want to represent it in metric and gauge potential form, we have to use again the definitions (\ref{def-Phi-Er})-(\ref{h-e}) to get the new functions\footnote{The $\gamma$ function is not affected by the Ehlers and the Harrison transformations, i.e. $\bar{\gamma} = \gamma$. Also $\D_x, \D_r, \Omega, \rho$ remain the same as the seed ones, as in eqs. (\ref{f-seed})-(\ref{w-seed}). The $\bar{\omega}=0$, in this simple case; for a more general solution including rotation see section \ref{sec:BH-in-Bertotti-Melvin-Swirling}.} ($\bar{f}, \bar{\omega} $) for the (\ref{LWP-spherical}) metric. Hence the final solution can be written as
\beq \label{LWP-spherical-new}
       ds^2 =-\bar{f} \, \D_\varphi^2 d\varphi^2 + \frac{1}{\bar{f}} \left[ \rho^2 dt^2 - e^{2\gamma}  \left( \frac{dr^2}{\Delta_r} + \frac{d x^2}{\Delta_x} \right) \right] \ ,
\eeq
with
\beq \label{fbar}
           \bar{f}(r,x) = \frac{- 4 B^4 r^2 \text{$\Delta_x$}}{\left[b (b+2 B) \left(B^2 m r x^2+1\right)-\left(b^2+2 b B+2 B^2\right) \Omega \right]^2} \ .
\eeq
The only non-null component of the electromagnetic potential, supporting the above metric, is given by
\beq \label{A-new-magn}
           \bar{A}_\varphi(r,x) = \frac{-2 r^2 (b+B) \text{$\Delta_x$}}{2 \Omega \left(B^2 m r x^2+1\right)+b r^2 (b+2 B) \text{$\Delta_x$}+2 B^2 r \left[x^2 \left(B^2 m^2 r+2m-r\right)+r\right]+2} \ .
\eeq
This metric describes a Schwarzschild black hole embedded into a combination of the Levi-Civita-Bertotti-Robinson and the Bonnor-Melvin external magnetic fields. All the limits are straightforward: for $b=0$ we return to the seed metric (\ref{schwarschil-br-spherical}), i.e. Schwarzschild-Bertotti-Robinson, because the Harrison transformation, in that case, is just the identity map. From the potential (\ref{A-new-magn}) we can deduce also that the value $b=-2B$ removes the Bonnor-Melvin contribution, so we again remain with the Schwarzschild-Bertotti-Robinson black hole.  For $B=0$ we have the Schwarzschild black hole embedded in the Bonnor-Melvin electromagnetic field \cite{ernst-magnetic}
\beq
          ds^2 = \left[1+\frac{b^2 r^2}{4} (1-x^2)\right]^2 \left[- \left(1- \frac{2m}{r} \right) dt^2 + \frac{dr^2}{1-\frac{2m}{r}} +\frac{r^2 dx^2}{(1-x^2)}\right] + \frac{r^2(1-x^2)}{\left[1+\frac{b^2 r^2}{4} (1-x^2)\right]^2} d\varphi^2 \ ,
\eeq 
\beq
           A_\m = \left[ 0,0,0, \frac{2br^2(1-x^2)}{4+b^2r^2(1-x^2)} \right] \ .
\eeq
On the other hand for $m=0$ we have the electromagnetic background, that is a Bertotti-Robinson-Bonnor-Melvin magnetic field, see section \ref{sec:Bertotti-Melvin}.\\  
Therefore the interpretation of the parameters of the solution is quite clear: the $b$ and $B$ parameters are related to the intensity of the Bonnor-Melvin and Bertotti-Robinson external magnetic field, while $m$ is related to the mass of the black hole.
The metric (\ref{LWP-spherical-new})-(\ref{fbar}) is of general type I, according to the Petrov classification, while the seed was of type $D$. This is expected because also the subcase $B=0$, the Schwarzschild-Bonnor-Melvin black hole, is of type I.\\
Notice that the above electrovacuum solution becomes a pure vacuum solution of Einstein theory (governed by field equations $R_{\m\n}=0$) for $b=-B$. In that case, we get a static generalisation of the Schwarzschild black hole, which is retrieved when the integration constant $B$ vanishes. The Petrov type is still I. No-hair theorems are circumvented here, because of the non-flat asymptotic and non-spherical symmetry. Further details about this vacuum solution will be studied elsewhere \cite{static-typeI-bh}. The swirling extension of this pure vacuum metric can be found in section \ref{sec:BH-in-Bertotti-Melvin-Swirling}, when $w=1$. From a mathematical point of view this vacuum solution is possible because the gravitational and electromagnetic seed Ernst's potentials are proportional as can be seen in eq. (\ref{Er-Phi}), therefore from the transformation (\ref{harrison}) it is clear that the electromagnetic field can be switched off by fine tuning the parameter of the Harrison transformation $\a$. A swirling generalisation of this vacuum spacetime can be deduced from the metric (\ref{fbar-general}) - (\ref{LWP-spherical-bar}) for $w=\pm1$ or in appendix \ref{app:swirling-bmbr-bh}. \\
To avoid conical singularities we have to set
\beq \label{Delta-phi}
          \D_\varphi = \frac{1}{1+B^2m^2} \ .
\eeq  
This is the same value of the seed, indeed the extra magnetic field is playing no role in the conical defect, in this simple case. However in rotating generalisations, as can be appreciated by the dependence of $\D_\varphi$ on $b$ in eq (\ref{D-phi-general}) the Bonnor-Melvin background expressly deforms the axial geometry of the spacetime.  \\
Curvature singularities are at $r=0$, since the Kretschmann scalar invariant diverges only there. In fact, possible divergences of the quadratic scalar invariants are, apart from $r=0$, at
\beq
             (b^2+2bB+2B^2)\sqrt{1+B^2r[r+(2m+B^2m^2r-r)x^2]} - b(b+2B)(1+B^2mrx^2)\ ,
\eeq
which is always positive in the range of the spherical coordinates we are using ($r \geq 0, \ -1\leq x \leq 1$), for any values of the physical parameters.\\
 The event horizon is located at
\beq
             r_h = \frac{2m}{1-m^2B^2} \ ,
\eeq
so it is not directly affected by the presence of the Bonnor-Melvin field, but only by the Bertotti-Robinson parameter. In any case, in order to have a proper defined event horizon the parameters have a limited range; for instance $|B| < 1/m$, with  $m>0$. When $B\to 0$ we recover the usual position of the Schwarzschild and Schwarzschild-Bonnor-Melvin black hole, $r=2m$. However, notice that the shape of the Schwarzschild-Bonnor-Melvin black hole is not spherical, but prolate, even though the horizon area remains $16 \pi m^2$, as the spherical black hole. For this Schwarzschild-Bertotti-Robinson-Bonnor-Melvin solution the geometry becomes more or less oblate, depending on the relative intensity of the external electromagnetic fields, as it can be appreciated looking at figure \ref{fig:picture-horizons}.

We can infer how the external magnetic fields deform the  shape of the event horizon, by directly computing the equatorial 
\beq
        C_e \ = \ \int_0^{2\pi} \sqrt{g_{\varphi\varphi}} \; d\varphi \ = \  \frac{4m\pi}{(1+m^2B^2)[1+m^2(b+B)^2]}   \ ,
\eeq
and the polar
\beq
        C_p \ = \ 2 \int_{-1}^{1} \sqrt{g_{xx}} \; dx \ = \ 8m \frac{(b+B)^2(1+m^2B^2)EllipticK(-m^2B^2)-b(b+2B)EllipticE(-m^2B^2)}{B^2(1+m^2B^2)^2}
\eeq 
circumferences. We observe that $C_s > C_e$, where $C_s$ are the spherical Schwarzschild ones (for $B=b=0$), on the other hand $C_p$ can be bigger or smaller than $C_s$ and $C_e$, depending on the relative values of the parameters $m,B,b$.  \\

\paragraph{Conserved charges and thermodynamics} The black hole temperature can be computed, as the surface gravity $\k_s$ on the event horizon as follows:
\beq
             T = \frac{\k_s}{2\pi} = \frac{(1+m^2B^2)^2}{8 \pi m} \ .
\eeq
The black hole entropy $S$ is taken as the Bekenstein-Hawking one, as a quarter of the event horizon area $A$
\beq
        S = \frac{A}{4} = \frac{1}{4} \int_0^{2\pi} d\varphi \int_{-1}^{1} \sqrt{g_{xx}g_{\varphi\varphi}}\ dx = \frac{4 \pi m^2}{(1+B^2m^2)^3} \ .
\eeq
The electric and magnetic black hole charges are null, as can be checked by computing\footnote{$n$ and $\s$ are respectively two time-like and spacelike unitary vectors normal to the boundary of the spacelike surface $\S$ surrounding the black hole, whose chosen normalisation is $n_\m n^\m = -1$ and $\s_\m \s^\m = 1$.} respectively 
\bea
         Q &=& - \int_{\p \S} d\varphi \, dx \, \sqrt{g_{xx} \, g_{\varphi \varphi}} \, n_\n \, \s_\n \, F^{\m\n} = 0 \ ,\\
         P &=& \frac{1}{8\pi} \int_{\p\S_\infty} \, F_{\m\n} \, dx^\m \w dx^\n = 0 \ .
\eea
The Komar mass of the black hole is given by
\beq
        M = \frac{m}{1+B^2m^2} \ .
\eeq
These values are coherent with the one for black holes in the Bertotti-Robinson background \cite{kerr-bertotti}, since the dependence on the Bonnor-Melvin parameter $b$ is not present. Actually it is known that the Bonnor-Melvin electromagnetic field does not modify the Schwarzschild quantities it surrounds: the mass, the area nor the temperature. So the Smarr law 
\beq \label{smarr}
              M = 2 T S \ ,
\eeq
can easily be checked. However, at first sight, the first law of black thermodynamics is not fulfilled by these values. This issue can be fixed by a proper normalisation of the time coordinate. In fact when the asymptotic is not flat, there is not a preferred way to define the clock flow of the observer. In practice we can find a suitable dilatation of the timelike coordinate $t$ by a constant factor $\a=\a(m,B,b)$ which depends on the integrating constant of the solution. This integrating factor has proven to be useful in fulfilling the first law in several black hole cases with a non constant curvature background such as Kerr-Newman-Bonnor-Melvin or accelerating black holes, see for instance \cite{kerr-melvin-mass} or \cite{thermo-acc}. As explained in detail in \cite{kerr-melvin-mass}, the integrating factor enters in the new definition of mass and temperature, but not into the horizon area, which respectively become  
\beq
     \bar{M} = \a M \quad , \hspace{2.3cm} \bar{T} = \a T \ .    
\eeq
Therefore the Smarr law (\ref{smarr}), after this time rescaling is still satisfied:
\beq
\bar{M}=2\bar{T}S \ .
\eeq
Moreover also the first law
\beq
                \d \bar{M} = \bar{T} \d S \ ,
\eeq
can be fulfilled, once we choose as an integrating factor\footnote{A similar procedure can be applied also to the case of the Kerr-Bertotti-Robinson solution. So, in that case, the first law can be satisfied taking into consideration also the variation of the constant in charge to remove the conical singularity.}
\beq
            \a = \frac{1}{\sqrt{1+m^2B^2}} \ .
\eeq
Note that the integrating factor is uniquely defined up to a multiplicative numerical factor, which we have settled to $1$
 to be in line with the standard asymptotically flat scenario, when $B=b=0$.\\
Furthermore is worth noting that the introduction of the integrating factor allows us to verify also the Cristodoulou-Ruffini formula \cite{ruffini}
\beq
        \bar{M} = \sqrt{\frac{S}{4\pi} +\frac{Q^2}{2}+\frac{\pi Q^4}{4S}} \ . 
\eeq
\\

\section{Bertotti-Robinson-Bonnor-Melvin electromagnetic background}
\label{sec:Bertotti-Melvin}

When the mass parameter $m$ vanishes from the metric (\ref{LWP-spherical-new})-(\ref{fbar}) we get just the external Bertotti-Robinson-Bonnor-Melvin magnetic field, which reads
\beq \label{back-metric}
 ds^2 = \L^2 \left[-(1+B^2r^2) dt^2 + \frac{dr^2}{1+B^2r^2} +\frac{r^2 dx^2}{(1-x^2)}\right] + \frac{r^2(1-x^2)}{\L^2 \ \big[1+B^2r^2(1-x^2)\big]^2} d\varphi^2 \ ,
\eeq
with 
\beq
           \L(r,x) = \frac{-b(b+2B)+(b^2+2bB+2B^2)\sqrt{1+B^2r^2(1-x^2)}}{2B^2[1+B^2r^2(1-x^2)]} \ .
\eeq
It is an electrovacuum regular metric of general relativity, whose magnetic potential is given by
\beq \label{back-A}
           A_\m = \left\{ 0,0,0, \frac{2(b+B)r^2(1-x^2)}{-(b^2+2bB+2B^2) r^2 (1-x^2)+2\Big[1+\sqrt{1+B^2r^2(1-x^2)}\Big]}  \right\} \ .
\eeq
The Kretschmann scalar invariant shows that the spacetime is everywhere free from curvature singularities, when we remove the black hole (i.e. for $m=0$). In fact, possible divergences of the scalar invariants could occur for
\beq
          (b^2+2bB+2B^2)\sqrt{1+B^2r^2(1-x^2)} - b^2 -2bB \ .
\eeq
But this quantity is always positive.\\
 Similarly the metric is also naturally free from conical singularities, for $m=0$, or equivalently from (\ref{Delta-phi}) we have $\D_\varphi=1$. According to the Petrov classification the background is of type D, as are both the Bertotti-Robinson and the Bonnor-Melvin spacetimes. However, in general, the conformal flatness is spoiled by the presence of the Bonnor-Melvin electromagnetic field. \\
 A similar background that combines Melvin and Bertotti-Robinson electromagnetic fields was studied in \cite{melvin-bertotti}, however, that can be singular in some portion of the parameter range, while the solution in (\ref{back-metric})-(\ref{back-A}) is not. \\
The interpretation of the background (\ref{back-metric})-(\ref{back-A}) is quite straightforward because when the $B$ parameter is zero (and $b\neq 0$) we recover the Bonnor-Melvin universe,
\beq
          ds^2 = \left[1+\frac{b^2 r^2}{4} (1-x^2)\right]^2 \left[- dt^2 + dr^2 +\frac{r^2 dx^2}{(1-x^2)}\right] + \frac{r^2(1-x^2)}{\left[1+\frac{b^2 r^2}{4} (1-x^2)\right]^2} d\varphi^2 \ ,
\eeq
with 
\beq
           A_\m = \left[ 0,0,0, \frac{2br^2(1-x^2)}{4+b^2r^2(1-x^2)} \right] \ .
\eeq
On the other had, when $b=0$ (or $b=-2B$) and $B\neq 0$ we get the Bertotti-Robinson electromagnetic background
\bea
          ds^2 &=& \frac{1}{1+B^2r^2(1-x^2)} \left[ -(1+B^2r^2)dt^2 + \frac{dr^2}{(1+B^2r^2)} + \frac{r^2 dx^2 }{(1-x^2)} + r^2(1-x^2) \ d\varphi^2 \right] \ , \\
          A_\m &=& \left[ 0, 0, 0, - \frac{Br^2(1-x^2)}{1+B^2r^2(1-x^2)+\sqrt{1+B^2r^2(1-x^2)}}  \right] \ .
\eea
This form of the metric has the advantage of possessing a clear limit to the flat Minkowski spacetime, for both $b=0$ and $B=0$.
However, thanks to the change of coordinates, suggested in \cite{kerr-bertotti}, i.e.
\beq
     r \to \frac{\sqrt{B^2 R^2 +\sin^2 \theta}}{B \cos \theta} \ , \qquad x \to \frac{BR}{\sqrt{B^2 R^2 +\sin^2 \theta}}  \ , \qquad B \to \frac{1}{e} \ , 
\eeq
the above metric takes the more familiar form
\beq
       ds^2 = - \left(1 + \frac{R^2}{e^2} \right) dt^2 + \frac{dR^2}{1 + \frac{R^2}{e^2}} + e^2 \left(d\theta^2 + \sin^2\theta \, d\varphi^2\right) \ .\\
\eeq  
In this case ($b=0$), the conformal flatness is restored, 
as certified by the scalar invariant $\Psi_2=0$.\\
A final notable example comes from the pure vacuum limit. Choosing $b=-B$ the electromagnetic field vanishes, so we remain with a pure general relativity background solution, without the Maxwell contribution in the energy-momentum tensor. This coupling suggests a deep relation between the Bertotti-Robinson and Bonnor-Melvin electromagnetic solutions. \\
To better identify the full background (\ref{back-metric})-(\ref{back-A}) we can perform the following change of coordinates
\bea \label{change-beginning}
       r &\to & \frac{\sqrt{4(p-1)[(b+B)^2-B^2p]+b^2(b+2B)^2q^2}}{b^2+2bB-2B^2(p-1)} \ , \\
        \qquad x &\to & \frac{b(b+2B)q}{\sqrt{4(p-1)[(b+B)^2-B^2p]+b^2(b+2B)^2q^2}} \ , \\
       \varphi & \to & \left( \frac{b}{2} + B \right) b \, \varphi \ ,
\label{change-end}
\eea
to arrange the metric (\ref{back-metric})-(\ref{back-A}) into the form 
\beq \label{16.27}
       p^2 \left[ - (1+B^2q^2) dt^2 + \frac{dq^2}{1+B^2q^2} + \frac{dp^2}{P(p)} \right] + \frac{P(p)}{p^2} d\varphi^2 \ ,
\eeq
with
\beq
          P(p) = -(b+B)^2 + \big[B^2+(b+B)^2\big] \, p - B^2 p^2 \ .
\eeq
The electromagnetic potential is transformed in
\beq \label{A-16.27}
           A_\m = \left[0, \, 0, \, 0 \, \frac{(p-1)(b+B)}{p} \right] \ .
\eeq
In this spacetime the two repeated principal null directions of the Weyl tensor are not expanding, thus, it belongs to type-D solutions of Kundt's class as described in section 16.4 of \cite{podolsky-book} or equivalently in section 2 of \cite{Podolsky:2018dpr}.\\ 
Also the swirling and the Bonnor-Melvin universe belong to this family of metrics represented in (\ref{16.27}). This observation is important to understand some key features of the background. In fact, as shown in \cite{swirling} and \cite{charging}, the swirling and the Bonnor-Melvin backgrounds correspond to the conjugation transformation, or equivalently the double Wick rotation ($t \to i \varphi , \ \varphi \to t$), of the Taub-NUT and the Reissner-Nordstrom (RN) spacetimes respectively, but where the base manifold does not have spherical symmetry, it is flat (or a cylinder whether the azimuthal coordinate is properly identified). Moreover, the two physical parameters that characterise the conjugated black holes are not free, but they are mutually related.  All that considered, it is natural to enquire about a possible link between the Bertotti-Robinson or the Bertotti-Robinson-Bonnor-Melvin solution in (\ref{back-metric})-(\ref{back-A}) and a conjugated black hole spacetime. At this scope we consider the topological Reissner-Nordstrom black hole modelled by the metric 
\beq \label{RN-topo}
        ds^2 = - \left( k - \frac{2m}{r} + \frac{Q^2}{r^2} \right) dt^2 + \frac{dr^2}{k - \frac{2m}{r} + \frac{Q^2}{r^2}} + \frac{dx^2}{1-kx^2} + r^2 (1-kx^2)d\varphi^2 \ ,
\eeq      
and whose electric field stems (up to an additive gauge constant)  from the potential 1-form
\beq \label{RN-A}
      A = A_\m dx^\m = -\frac{Q}{r} \, dt\ . 
\eeq
The parameter $k$ determines the topology of the black hole base manifold: when $k$ is positive, zero or negative we have a spherical, flat or hyperbolic symmetry and in principle $k$ can be normalised to $1, 0$ or $-1$, without loss of generality. In this case, the angular coordinate $x$, depending on the values of $k$, can have a different ranges with respect to the usual spherical interpretation, where, being $x=\cos \theta$ it is interpreted as a standard polar angle with domain $[-1,1]$. For instance, when $k=0$, the range of the $x$ coordinate is $[-\infty,\infty]$.  \\
Remarkably the background, written as in eqs. (\ref{16.27})-(\ref{A-16.27}), can be retrieved from the above topological charged black hole solution by the double Wick rotation of the Killing coordinates ($t,\ \varphi$) and the following redefinition of the physical parameters and coordinates
\beq
           t \to i \varphi \ , \qquad \varphi \to i t \ , \qquad Q \to i (b+B) \ ,  \qquad k \to -B^2 \ , \qquad m \to  -\frac{b^2}{2} -bB-B^2 \ , \qquad r \to p \ , \qquad x \to q \ .
\eeq 
When dealing with this kind of analytic continuations, an imaginary rotation of the gauge potential constants (in this case $Q$), is needed to remain with a real potential and well defined energy conditions for the Maxwell energy momentum tensor. \\
Note that the double Wick rotated Reissner-Nordstrom metric is, in principle, more general with respect to the Bonnor-Melvin-Bertotti-Robinson background, because the first carries three independent physical parameters instead of the two of the second metric.\\ 
As a consequence of the mapping between the metrics (\ref{back-metric}) and (\ref{RN-topo}), which is apparent also when the background is written in the coordinates ($t,p,q,\varphi$), the Bertotti-Robinson and the Bonnor-Melvin electromagnetic fields are explicitly proportional. This basically occurs for the same reason why the electromagnetic potential of the Reissner-Nordstrom black hole is not depending on the base manifold topology, thus, eq. (\ref{RN-A}) is independent of the parameter $k$. This is the main point that allows us to fine tune the two intensities of the respective electromagnetic fields ($b$ and $B$) to erase completely the Maxwell energy-momentum tensor and reduce the black hole into the Bonnor-Melvin-Bertotti-Robinson electromagnetic universe, as described in (\ref{LWP-spherical-new})-(\ref{A-new-magn}), to a pure vacuum solution. Hence the role of $B$ is double, it is not only related to the intensity of the Bertotti-Robinson electromagnetic field, but also on the geometry of the solution: considering the double Wick rotated dual RN solution, $B$ determines both the intensity of the gauge field and the hyperbolic topology of the ($x-\varphi$) section of the solution, through $Q$ and $k$ respectively. On the other hand, this section appears as the $AdS_2$ in the metric (\ref{16.27}) for constant $p$ and $\varphi$. It's also evident that, when $B$ is null, $k$ switches from negative to zero, determining the flat sector of the Bonnor-Melvin geometry.   \\  
The third and last possible case refers to positive values of $k$ in the double Wick rotated geometry of the topological Reissner-Nordstrom solution (\ref{RN-topo})-(\ref{RN-A}). Physically this possibility represents a charged generalisation of the Witten's bubble of nothing.
\\
Hence, both the Bonnor-Melvin and the Bertotti-Robinson backgrounds represent fundamentally the same electromagnetic field, but immersed in geometries with different symmetry properties. Then it is not surprising that the spherical black hole embedded into a geometric background, which shares with it the same symmetry properties, are algebraic of the same type, such as the Schwarzschild black hole immersed into the Bertotti-Robinson electromagnetic field. On the other hand the Schwarzschild spacetime embedded into the Bonnor-Melvin universe, which has cylindrical symmetry, have a more general Petrov algebraic type ($I$). So, when the symmetries of the background and of the black hole are coherent, the solution can be of special type. This compatibility improves the alignment between principal null directions of the metric and of the electromagnetic field, assuring a special algebraic type, such as $D$. \\
In any case, since the change of coordinates (\ref{change-beginning})-(\ref{change-end}) is not everywhere invertible, the nature of the $B$ and $b$ parameters may differ with respect to the background originally written in (\ref{back-metric})-(\ref{back-A}). For instance, in the latter representation of the background, the conformal flatness of the metric is clear when the Bonnor-Melvin contribution vanishes, for $b \to 0$ or $b \to -2B$; while this behaviour is not immediately apparent in the representation (\ref{16.27})-(\ref{A-16.27}).\\

Finally note that when the background is written in the form (\ref{16.27})-(\ref{A-16.27}) it admits an easy generalisation to the presence of the cosmological constant, just by upgrading the $P$ function to
\beq
                P(p) = -(b+B)^2 + \big[B^2+(b+B)^2+c\big] \, p - B^2 p^2 - \frac{\Lambda}{3} p^4 \ .
\eeq
The new integration constant $c$ can be set to zero to recover exactly the background solution in eqs. (\ref{back-metric})-(\ref{back-A}), in the vanishing $\L$ limit.
\\

\section{Further generalisation: black hole in a swirling Levi-Civita-Bertotti-Robinson-Bonnor-Melvin electromagnetic background}
\label{sec:BH-in-Bertotti-Melvin-Swirling}

There are several possible generalisations of the spacetimes described by eqs. (\ref{LWP-spherical-new})-(\ref{fbar}). For instance using a complex parameter for the Harrison transformation, thus including an electric Bonnor-Melvin background or considering a seed with both external Bertotti-Robinson electric and magnetic field. Furthermore a more general Lie-point symmetry can be used, instead of the Harrison we exploited in section \ref{sec:generation}. We can compose the Harrison transformation with the Ehlers transformation to add both an external Bonnor-Melvin electromagnetic field and a swirling gravitational background. The composition of the two continuous and one-parameter transformations is unique, because as shown in \cite{Type-I}, the two elements of the group of the Lie-point symmetries of the electromagnetic Ernst equations, $SU(2,1)$, commute\footnote{This is not a trivial property because, in general, the symmetry group of the Ernst equations is not abelian.}. Therefore the resulting solution that can be generated with the composition of these two maps is also unique\footnote{Obviously, the composition with other discrete symmetry transformations can add some further physical features, such as angular momentum, NUT or monopolar electromagnetic charges.}. This kind of composite map has already used to generate a number of different black holes solutions, see for instance \cite{illy} or \cite{dipinto}, however in this case, since the seed already carry the Bertotti-Robinson external electromagnetic field, the black hole solutions built in this paper have even a more general background.\\

As seed we consider a slightly more general spacetime with respect to the Schwarzschild in the Bertotti-Robinson magnetic field of section \ref{sec:generation}. We take its electric generalisation, which means, because of the electromagnetic duality in four-dimensions, that the metric remains the same as the one in section \ref{sec:generation}, described by eqs. (\ref{LWP-spherical})-(\ref{w-seed}), but the electromagnetic potential is rotated into
\beq
       A_\m = \left\{\frac{Bx(2m+B^2m^2r-r)\sqrt{1-w^2}}{\Omega(r,x)}\,, \ 0\,, \, 0\,, \ \frac{w \D_\varphi[1+B^2mrx^2-\Omega(r,x)]}{B\, \Omega(r,x)}\right\}
\eeq
The parameter $w \in [-1,1]$ defines the rotation of the magnetic field into the electric field: for $w=\pm1$ we have the purely magnetic seed solution (\ref{LWP-spherical})-(\ref{A-seed-magn}), while for $w=0$ we get a purely electric Bertotti-Robinson background. For $w \in (-1,0) \cup (0,1)$ we have a combination of both electric and magnetic external fields.\\
From the definitions (\ref{A-tilde-e})-(\ref{h-e}) we deduce, up to a gauge constant, the seed
\beq 
       \tilde{A}_t(r,x) =  \frac{-\sqrt{1-w^2}(1+B^2mrx^2)}{B\Omega(r,x)} \ ,
\eeq 
and 
\beq     
       h(r,x) =\frac{2w}{B} \, \tilde{A}_t(r,x) \ .
\eeq
Therefore the complex Ernst potentials for the seed metric read
\bea
            \Er(r,x) &=&  \frac{w^2(2+2B^2mrx^2-\Om)- \Om -2iw\sqrt{1-w^2} (1+B^2mrx^2)}{B^2\Om} \ , \\
            \mathbf{\Phi} (r,x) &=& \frac{w(1+B^2mrx^2-\Omega) -i\sqrt{1-w^2}(1+B^2mrx^2)}{B \, \Om}  \ .
\eea
The transformation of the seed solution into a new one is done due to the composition of the Harrison and Ehlers transformation \cite{Type-I}
\beq\label{harrison-ehlers}
      \Er \longrightarrow \bar{\Er} = \frac{\Er}{1-2\a^*\mathbf{\Phi} -\a\a^* \Er +i \jmath \Er} \qquad \quad ,  \quad  \qquad  \mathbf{\Phi} \longrightarrow  \bar{\mathbf{\Phi}} = \frac{\mathbf{\Phi} + \a \Er}{1-2\a^*\mathbf{\Phi} -\a\a^* \Er + i \jmath \Er} \quad .        
\eeq
This map is a three (real) parameter symmetry of the complex Ernst equations, however we choose $\a=b/2$ with\footnote{This choice means that we are writing, in the metric form, only the magnetic contribution of the Bonnor-Melvin field.} $b \in \mathbb{R} $. When $\jmath=0$ we recover exactly the Harrison map (\ref{harrison}), while when $\a=0$ we have the Ehlers one. To write the new solution into metric and vector potential form ($\bar{g}_{\m\n}, \bar{A}_\m$) we have to use again (\ref{def-Phi-Er}) - (\ref{h-e}); hence we get
\bea \label{fbar-general}
      \bar{f}(r,x) &=& \frac{f(r,x)}{|1-2\a^*\mathbf{\Phi} -\a\a^* \Er +i \jmath \Er|^2} \ , \\
      \bar{\om}(r,x)    &=& \frac{x(r-2m-B^2m^2r)}{4B^2\Om} \bigg\{ 8B\jmath w[2Bw-b(1-w^2)]  \label{omega-gen}\\
       &+&\sqrt{1-w^2}[8bB^3+12b^2B^2w+w(1+w^2)(16\jmath^2+b^4)+2b^3B(1+3w^2)] \bigg\}  \ , \nn \\
        \bar{A}_\varphi(r,x)   &=&   \textsf{Re} (\bar{\mathbf{\Phi}})\, \D_\varphi \ , \\ 
      \bar{A}_t(r,x)   &=&  - \bar{\om}\, \textsf{Re} (\bar{\mathbf{\Phi}}) -\frac{x(r-2m-B^2m^2r)\sqrt{1-w^2}}{4B^2\Om} \bigg\{12bB^2w+3b^2B(1+w^2)  + \\
       & & \hspace{2cm} +2b^3w(1+w^2) +  4B(B^2-\jmath w\sqrt{1-w^2}) \bigg\} \ .
\eea
The final solution\footnote{A Mathematica notebook containing the explicit solution can be found in between the source arXiv files.} is obtained by inserting the above values into the LWP metric
\beq \label{LWP-spherical-bar}
       \bar{ds}^2 = -\bar{f} ( \D_\varphi d\varphi - \bar{\omega} dt)^2 + \bar{f}^{-1} \left[ \rho^2 dt^2 - e^{2\gamma}  \left( \frac{dr^2}{\Delta_r} + \frac{d x^2}{\Delta_x} \right) \right] \ .
\eeq
On the other hand the $\g, \D_x, \D_r, \rho$ functions remain unchanged by the Harrison transformation, so it remains the same of the seed, as in (\ref{f-seed})-(\ref{w-seed}). \\
This spacetime models a Schwarzschild black hole in an external gravitational rotating background and an external electromagnetic field which is the combination of both the Bertotti-Robinson and the Bonnor-Melvin backgrounds.\\ 
A simpler solution, describing the a Schwarzschild black hole immersed into the swirling Bonnor-Melvin-Bertotti-Robinson background, for a $w=\pm1$ can be found in the appendix \ref{app:swirling-bmbr-bh}; whereas the spacetime presented in section \ref{sec:generation} is retrieved\footnote{To get the proper limits for specific values of the parameters, in some cases, it can be necessary to use the gauge freedom in rescaling the Killing coordinates or to adjust the additive gauge constants in $\bar{\omega}$ and $\gamma$.} eliminating the swirling contribution by also setting  $\jmath=0$. The metric is stationary rotating because of two independent effects: the tidal drag of the swirling universe and because of the interaction between the Bertotti-Robinson electric and Bonnor-Melvin magnetic field, that is for $w^2\neq1, b\neq 0$ and $B\neq 0$. \\
All conical singularities are avoided for
\beq \label{D-phi-general}
      \D_\varphi = \frac{8b^2B(3B+2bw)(1-w^2)-32bB\jmath(1-w^2)^{\frac{3}{2}}+(b^4+16\jmath^2)(1+2w^2-3w^4)+16B^2(B^2+4\jmath w\sqrt{1-w^2})}{16B^4(1+B^2m^2)}
\eeq
The spacetime is naturally free from torsion singularities or Misner strings.\\
Ergoregions that extend to infinity are expected, for some values of the parameters range, because both the black holes in the Bonnor-Melvin and swirling backgrounds have them. For $\jmath=0$ and $b=0$ simultaneously the metric is diagonal and there are not ergoregions at infinity. The metric can be diagonal also for some other specific values of the constants in the brackets of eq. (\ref{omega-gen}).   \\
The black hole can easily be removed from the spacetime just setting $m=0$; in that case we remain with a {\it swirling-Bertotti-Robinson-Bonnor-Melvin background}, which is the swirling and electric generalisation of the one of section \ref{sec:Bertotti-Melvin}. Also in this case the background is of type D, otherwise when the black hole is present ($m\neq0$) the spacetime becomes of more general type $I$. This is an analogous behaviour with respect to the Schwarzschild-Bonnor-Melvin subcase. 
When the mass and Bertotti-Robinson electromagnetic field is null ($m=0, B=0$) one recovers the swirling Bonnor-Melvin background found by Harrison \cite{harrison}. As seen in the previous section the Bonnor-Melvin-Bertotti-Robinson corresponds to the double Wick rotation of the topological Reissner-Nordstrom black hole, moreover the swirling background is the double Wick rotation of the Taub-NUT solution \cite{swirling}. Therefore it is clear that the swirling-Bertotti-Robinson-Bonnor-Melvin background stems from the double Wick rotation of the topological Reissner-Nordstrom-NUT black hole solution.   \\
Finally note that when $w=\pm1$ the seed electromagnetic and gravitational Ernst potentials are proportional, therefore, also after the Ehlers-Harrison transformation (\ref{harrison-ehlers}) the Ernst electromagnetic potential can be vanished by a proper choice of the $\a$ parameter. Therefore tacking $2\a=b=-B$ we obtain a pure vacuum solution representing a swirling black hole into an external gravitational background, a stationary generalisation of \cite{static-typeI-bh}. The metric can be written as (\ref{LWP-spherical-bar}) with 
\bea \label{swirling-bh-hyberbolic-bubble}
           \bar{f}(r,x) &=& \frac{-4 B^4 r^2(1-x^2)(1+B^2m^2r^2) }{16 \jmath^2 (1+B^2mrx^2-\Omega)^2+B^4(1+B^2mrx^2+\Omega)^2}  \ , \\
           \bar{\omega}(r,x) &=& \frac{4 \jmath x (r-2m-2B^2m^2r)}{\Omega}  + \om_0 \ .\label{swirling-bh-hyberbolic-bubble-omega}
\eea
The gravitational background represents a swirling generalisation of the hyperbolic variant\footnote{To be more precise, we recall that the bubble of nothing corresponds to the double Wick rotation of the Schwarzschild black hole (with a standard spherical horizon). On the other hand this background can be thought as the double Wick rotation of a Schwarzschild metric with an hyperbolic base manifold. } of the metric of the expanding bubble of nothing \cite{witten}, however note that, in this hyperbolic case, there are no bubble horizons. Thus, thanks to the analytic continuation of the parameter $B \to i \bar{B}$ in (\ref{swirling-bh-hyberbolic-bubble})-(\ref{swirling-bh-hyberbolic-bubble-omega}), we obtain a swirling extension of the Schwarzschild black hole inside a Witten's bubble \cite{bh+bubble}.
 \\
Similarly, when there is no swirling contribution, that is for $\jmath=0$, another vacuum black hole metric can be obtained from the solution (\ref{fbar-general})-(\ref{LWP-spherical-bar}), for $4B^2+4bBw-b^2(1-w^2)=0 $. 

In the figure \ref{fig:grafico} the structure of the new solutions presented in this article and their relations with the known spacetimes are summarised. \\

\tikzstyle{rect} = [draw,rectangle, fill=white!20, text width =3cm, text centered, minimum height = 1.5cm,scale=0.9]
\begin{figure}[H]
\vspace{0.3cm}

\begin{center}

\begin{tikzpicture}

\hspace{-0.2cm}    \node[rect,scale=1.2,label=above :{type I \quad \qquad \qquad (\ref{fbar-general})-(\ref{LWP-spherical-bar})},text width=4.5cm,node distance=1.5cm,line width=1.3pt](PD double nut){\bf Schwarzschild-Bertotti-Robinson-Bonnor-Melvin in a swirling background \\ ($m,B,w,b,\jmath$)};
    \node[rect,scale=1.2,anchor=north,label=above :{type I \hspace{2.4cm} \cite{illy}},
    text width=4.5cm,below of=PD double nut,node distance=4.3cm](acc kerr newman nut){Schwarzschild-Bonnor-Melvin in swirling background\\ ($m,b,\jmath$)};
    \node[rect,scale=1.1,anchor=north,label=above left :{type I },text width=4.5cm, left of= acc kerr newman nut, node distance=6cm,line width=1.3pt](PD-NUTs){\bf Schwarzschild-Bertotti-Robinson-Bonnor-Melvin \\ ($m,B,w,b$)};\node[rect,scale=1.1,anchor=north,label=above :{type I \quad \quad  (\ref{LWP-spherical-new})-(\ref{fbar})},text width=4.2cm, above of= PD-NUTs, node distance=4cm,line width=1.3pt](PD-NUTs-sopra){\bf Schwarzschild-Bertotti-Robinson-Bonnor-Melvin \\ ($m,B,b$)};
    \node[rect,scale=1.1,anchor=north,label=above :{\qquad type I},text width=4.5cm, right of= acc kerr newman nut, node distance=6cm,line width=1.3pt](PD){\bf Schwarzschild-Bertotti-Robinson in swirling background\\ ($m,B,w,\jmath$)};
    \node[rect,scale=1.2,anchor=north,label=above :{type I \qquad \qquad \qquad \qquad  \cite{ernst-magnetic}},text width=4.3cm, below of=PD-NUTs, node distance=4cm](acc RN NUT){Schwarzschild-Bonnor-Melvin ($m,b$)};
    \node[rect,scale=1.2,anchor=north,label=above :{type I \hspace{2.5cm} \cite{swirling}},text width=4.5cm, right of= acc RN NUT,node distance=5.6cm](kerr newman nut){Schwarzschild in swirling background ($m, \jmath$)};
  \node[rect,scale=1.2,anchor=north,label=above :{type D \hspace{2.5cm} \cite{kerr-bertotti}},text width=4.3cm, below of= PD, node distance=4cm](mBw){Schwarzschild-Bertotti-Robinson ($m, B, w$)};   
    \node[rect,scale=1.2,anchor=north,below of=acc RN NUT,node distance=3.2cm,label=above:{\; type D \hspace{2.1cm} \cite{bonnor}-\cite{melvin}},text width=4.3cm](RN NUT){Bonnor-Melvin universe ($b$)};
    \node[rect,scale=1.2,anchor=north,below of=kerr newman nut,node distance=3.2cm, label=above :{type D \qquad \qquad \qquad },text width=4.4cm](RN charged RIndler){Schwarzschild black hole ($m$)};
      \node[rect,scale=1.2,anchor=north,below of=mBw,node distance=3.2cm, label=above :{type D \hspace{2cm} \cite{levi-civita}-\cite{robinson} },text width=4.3cm](Bw){Bertotti-Robinson ($B,w$)};

 \node[rect,scale=1.2,anchor=north,line width=1.3pt, above of=PD,node distance=3.4cm, label=above :{type D \qquad \qquad \qquad },text width=4.1cm](Bwbj){\bf Bertotti-Robinson-Bonnor-Melvin in swirling background \\($B,w,b,\jmath$)};

\draw[->] (PD-NUTs) -- node {\hspace{-1.1cm}$w=1$}(PD-NUTs-sopra);
\draw[->] (acc kerr newman nut) -- node [right] {$\jmath$ = 0} (acc RN NUT);
\draw[->] (PD double nut) -- node {$B=0$}(acc kerr newman nut);
\draw[->] (PD double nut) -- node [right] {\; \ $b$ = 0}(PD);
\draw[->] (PD double nut) -- node [left] {\; $\jmath$ = 0}(PD-NUTs);
\draw[->] (acc kerr newman nut) -- node [right] {\; \; $b$ = $0$}(kerr newman nut);
\draw[->] (PD-NUTs) -- node [right] {\; $B=0$}(acc RN NUT);
\draw[->] (kerr newman nut) -- node [ right] {$\jmath$ = 0} (RN charged RIndler);
\draw[->] (acc RN NUT) -- node [right] {\; \; $m$ = 0} (RN NUT);
\draw[->] (PD) -- node [right] {\; \; $\jmath$ = 0} (mBw);
\draw[->] (PD) -- node [right] {\; \; $B$ = 0} (kerr newman nut);
\draw[->] (acc RN NUT) -- node [right] {\; \; $b$ = 0} (RN charged RIndler);
\draw[->] (mBw) -- node [right] {\; \; $B$ = 0} (RN charged RIndler);
\draw[->] (mBw) -- node [right] {\; \; $m$ = 0} (Bw);
\draw[->] (PD double nut) -- node [above] {$m$ = 0} (Bwbj);
\end{tikzpicture}
\vspace{0.1cm}
\caption{Map of the solutions of the Einstein-Maxwell theory presented in this article. The novel spacetimes, not yet known in the literature, are emphasized in bold line rectangles. The new Schwarzschild-like solutions belong to the Petrov Type I. The more general black hole (\ref{fbar-general})-(\ref{LWP-spherical-bar}) carries, apart from the mass $m$ and the intensity of the swirling background $\jmath$,  a couple of integrating constants related to the Bertotti-Robinson field ($B$ and $w$) and another one ($b$) for the Melvin-Bonnor field.}
\label{fig:grafico}
\end{center}
\end{figure}
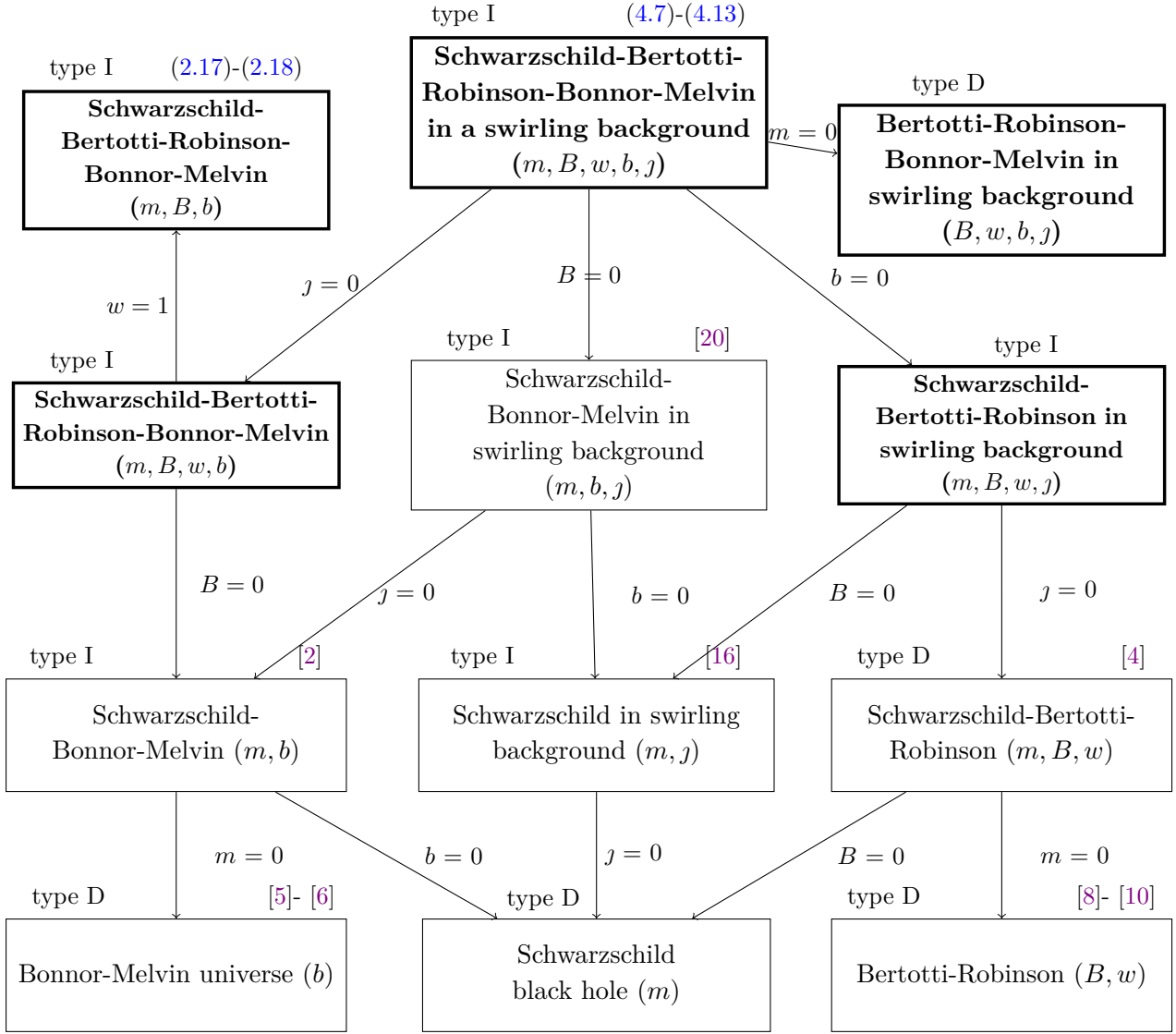

\newpage

\section{Conclusions}

In this article we built an analytical and exact black hole metric in general relativity coupled with the Maxwell electromagnetic field. We first used a Lie-point symmetry of the Ernst equations, the Harrison transformations. In this way we were able to extend and merge, at the same time, two families of black holes embedded in an external electromagnetic field, the ones in the Bertotti-Robinson and the ones in the Bonnor-Melvin electromagnetic backgrounds. The resulting spacetimes are naturally of Petrov type $I$, as their known subcases, the black holes in the Bonnor-Melvin electromagnetic fields.\\
This general result can improve the phenomenology of both previous systems, at least in some range of the parametric space or as a local model, for a certain range of the radial distance from the collapsed star.\\
We focused mainly on a static spacetime for simplicity, however, we have shown that also stationary extensions of the above spacetime can be readily built, just considering a complete electromagnetic Schwarzschild-Bertotti-Robinson seed, or by composing the Harrison with the Ehlers transformation, thus adding an extra integrating constant related to the intensity of the rotation of the external gravitational field. All the limits to the previous known metrics, included in the new solution, are clear and well defined. \\ 
The analysis of the background allows us to understand the relation between the two homogeneous electromagnetic fields in general relativity: the Bonnor-Melvin and the Bertotti-Robinson solutions. They basically represent the same electromagnetic potential, but immersed in geometries with different symmetries: cylindrical the first, while spherical the latter.\\
Further generalisations of the presented family of solutions are possible by operating with a complex Lie-point parameter, instead of a real one, in the Harrison transformation; or starting with a more general seed, which may include angular momentum, acceleration, gravitomagnetic or electromagnetic monopolar charges, such as the recent type-D metrics presented in \cite{kerr-newman-bertotti}. Also generalisation to some class of scalar tensor theories or ModMax coupling are possible, since the Harrison and the Ehlers transformations have been adapted in these settings in \cite{marcoa-embedding}, \cite{herdeiro}. On the other hand, extension in the presence of the cosmological constants, apart from the background spacetime for $m=0$, are more difficult since the generating techniques are not known to work in this setting. That's because some symmetries of the field equations seem to be broken\footnote{Nevertheless the Bonnor-Melvin solution can be generalised in the presence of the cosmological constant \cite{charging}.} when the cosmological constant is non-zero \cite{charging}.\\
A specific interesting subcase can be obtained by fine tuning the intensity of the two external electromagnetic fields to remove completely the electromagnetic contribution. Hence we remain with a type $I$, static black hole in pure general relativity, which extends the Schwarzschild solution by the presence of a continuous gravitational hair, for more details see \cite{static-typeI-bh}. Swirling generalisation are also naturally included in the metric, depending on the reality or imaginary nature of the  parameter $B$; these spacetimes describe a black hole inside a rotating expanding bubble of nothing, in the former case, or its topological extension, in the latter case.  \\

\vspace{2mm}

\paragraph{Acknowledgements}
{\small I would like to thank Marcello Ortaggio, Jiri Podolsky, Hryhorii Ovcharenko and Andrea Di Pinto for interesting correspondence on this subject.
A Mathematica notebook containing the main solutions presented in this article can be found in the arXiv source folder.}\\

\newpage

\appendix

\section{More embedding pictures of the event horizon}
\label{app:more-figures}

\begin{figure}[h!]
\captionsetup[subfigure]{labelformat=empty}
\centering
\hspace{-1cm}
\subfloat[\hspace{0.1cm} $m=1$, $B=0.1$, $b=0.3$]{{\includegraphics[scale=0.45]{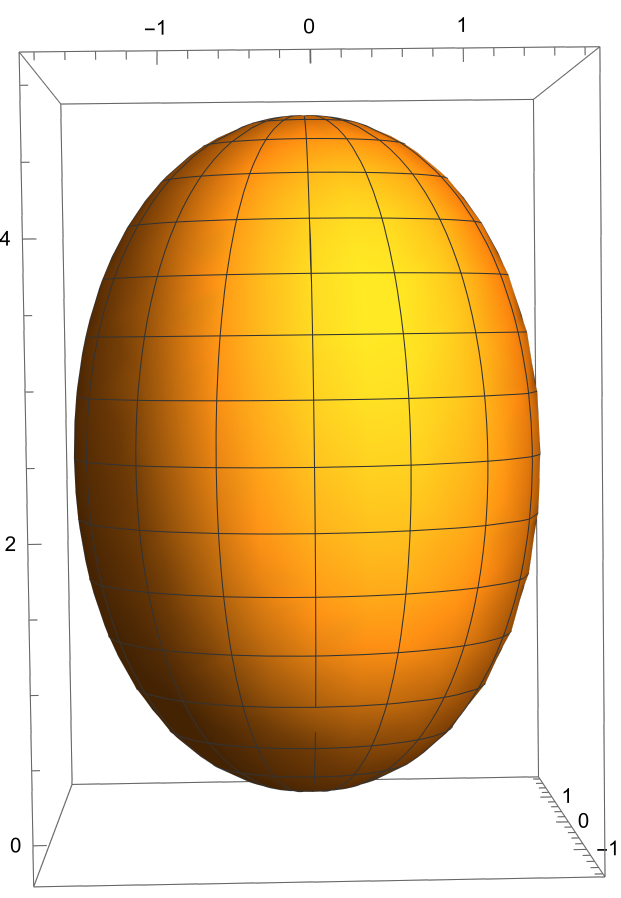}}}%
\subfloat[\hspace{0.8cm} $m=1$, $B=2$, $b=1$]{{ \hspace{0.5cm} \includegraphics[scale=0.6]{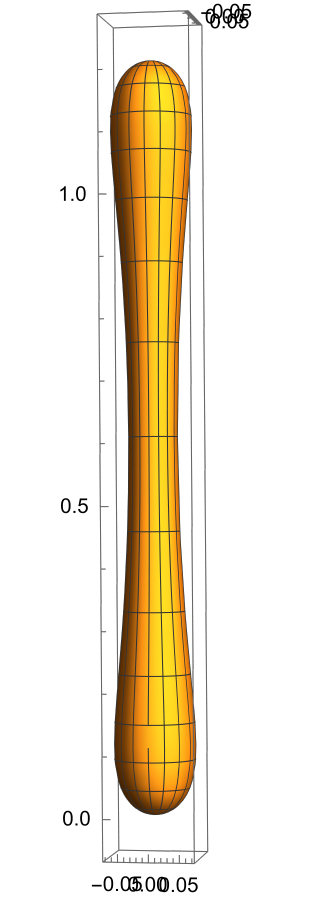}}}%
\subfloat[\hspace{0.8cm} $m=1$, $B=15$, $b=0.1$]{{\hspace{1cm}
\includegraphics[scale=0.7]{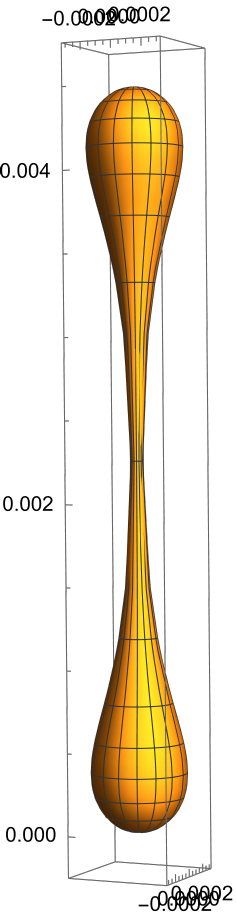}}}
\caption{\small More embeddings in Euclidean three-dimensional space $\mathbb{E}^3$ of the event horizon of the black hole described by metric (\ref{LWP-spherical-new})-(\ref{A-new-magn}), distorted by the presence of both the external Bertotti-Robinson and Bonnor Melvin magnetic backgrounds, for different values of the parameters $b$ and $B$, and $m=1$.}%
\label{fig:picture-horizons-more}
\end{figure}

In figure \ref{fig:picture-horizons-more} some more pictures of the event horizon of the black hole solution of section \ref{sec:generation} are shown, for different intensities of the magnetic fields.

These picture naturally suggest that the presence of strong electromagnetic fields can stretch a single black hole to nucleate a couple of them.\\

\newpage

\section{Schwarzschild black hole embedded in a swirling\\ Bertotti-Robinson-Bonnor-Melvin background}
\label{app:swirling-bmbr-bh}

We present here an explicit generalisation, by considering a swirling effect to the background, of the solution built in section \ref{sec:generation}. This solution describes a Schwarzschild back hole embedded into a swirling Bertotti-Robinson-Bonnor-Melvin background. It is a subcase of the more general spacetime of section \ref{sec:BH-in-Bertotti-Melvin-Swirling}, in fact it can be obtained as a limit for $w\to1$ of (\ref{fbar-general})-(\ref{LWP-spherical-bar}). Alternatively it can be generated, thanks to the combination of the Ehlers and Harrison transformation (\ref{harrison-ehlers}), as described in section \ref{sec:BH-in-Bertotti-Melvin-Swirling}, but starting with a purely magnetic background  seed, i.e. with $w^2=1$.\\
The metric can be written as\footnote{A Mathematica notebook containing this solution can be found between the source files of the arXiv repository.} 
\beq \label{LWP-spherical-tilde-app}
       \tilde{ds}^2 = -\tilde{f} ( \tilde{\D}_\varphi d\varphi - \tilde{\omega} dt)^2 + \tilde{f}^{-1} \left[ \rho^2 dt^2 - e^{2\gamma}  \left( \frac{dr^2}{\Delta_r} + \frac{d x^2}{\Delta_x} \right) \right] \ ,
\eeq
with
\bea
      \tilde{f}(r,x) &=& \frac{-4B^4r^2 \D_x}{16\jmath^2(1+B^2mrx^2-\Om)^2 + [b(b+2B)(1+B^2mrx^2)-(b^2+2bB+2B^2)\Om)]^2}  \ , \\  
      \tilde{\om} (r,x) &=& \frac{4\jmath x[r-m(2+B^2mr)]}{\Om} + \om_0 \ .
\eea
The non-null component of the electromagnetic potential, up to an additive constant, are
\bea
       A_t(r,x) &=& A_{t_0} - \tilde{\om}(r,x) \tilde{A}_{\varphi}(r,x)\ , \\
       A_\varphi (r,x) &=& \frac{(b+B)[2bB^2(b+2B)r^2 \Delta_x+4(b+B)^2(1+B^2mrx^2-\Om)\Omega]}{16\jmath^2(1+B^2mrx^2-\Om)^2 + [b(b+2B)(1+B^2mrx^2)-(b^2+2bB+2B^2)\Om)]^2} \ .\label{Aphi-app}
\eea 
$\tilde{\D}_\varphi, \, \om_0, \, A_{t_0}$ are constant, while functions $\D_r, \, \D_x, \Om$ are defined in eqs. (\ref{f-seed}). Conical singularities are not present when $\tilde{\D}_\varphi$ takes the same value as in (\ref{Delta-phi}).

While both the solutions (\ref{LWP-spherical-tilde-app})-(\ref{Aphi-app}) and (\ref{fbar-general})-(\ref{LWP-spherical-bar}) describe a Schwarzschild in Bertotti-Robinson-Bonnor-Melvin background there is a significative difference: the former metric rotates only because the presence of the swirling background ($\tilde{\om}=0$ when $\jmath=0$), while in the latter metric the rotation also caused by the interaction between the Bertotti-Robinson contribution to the electric field and the Bonnor-Melvin magnetic field ($\bar{
\om} \neq 0$ for $\jmath=0$, unless $w^2=1$).\\
When $m=0$ we have the {\it swirling Bertotti-Robinson-Bonnor-Melvin} background, where the Bertotti-Robinson is only magnetic. A more general background, including the electric component of the Bertotti-Robinson field, is given in section \ref{sec:BH-in-Bertotti-Melvin-Swirling}.\\
Also in this case the Bonnor-Melvin electromagnetic field can erase the Bertotti-Robinson one by setting $b=-B$. We remain with a vacuum solution of general relativity describing a Schwarzschild black hole in a swirling gravitational background. The metric is given by (\ref{swirling-bh-hyberbolic-bubble})-(\ref{swirling-bh-hyberbolic-bubble-omega}). \\
Then, as already noted at the end of section \ref{sec:BH-in-Bertotti-Melvin-Swirling}, by further rotate the B parameter into the imaginary plane $B \to i \bar{B}$ the resulting vacuum metric describe instead a Schwarzschild black hole in an expanding swirling Witten bubble of nothing.


\newpage


\begin{thebibliography}{99}


\bibitem{Eatough:2013nva}
R.~P.~Eatough, H.~Falcke, R.~Karuppusamy, K.~J.~Lee, D.~J.~Champion, E.~F.~Keane, G.~Desvignes, D.~H.~F.~M.~Schnitzeler, L.~G.~Spitler and M.~Kramer, \textit{et al.}
{\it ``A strong magnetic field around the supermassive black hole at the centre of the Galaxy''},
\href{https://doi.org/10.1038/nature12499}{Nature \textbf{501} (2013), 391-394};
\href{https://arxiv.org/pdf/1308.3147}{\tt [arXiv:1308.3147 [astro-ph.GA]]}

\bibitem{ernst-magnetic} 
  F.~J.~Ernst,
   {\it ``Black holes in a magnetic universe''},
    \href{https://doi.org/10.1063/1.522781}{J.\ Math.\ Phys.\  {\bf 17}, no. 1, 54 (1976).}
    
\bibitem{garcia-alekseev}
G.~A.~Alekseev and A.~A.~Garcia,
{\it ``Schwarzschild black hole immersed in a homogeneous electromagnetic field''},
\href{https://doi.org/10.1103/PhysRevD.53.1853}{Phys. Rev. D \textbf{53} (1996), 1853-1867}.

\bibitem{kerr-bertotti}
J.~Podolsky and H.~Ovcharenko,
{\it ``Kerr black hole in a uniform magnetic field: An exact solution''}
\href{https://arxiv.org/pdf/2507.05199}{\tt [arXiv:2507.05199 [gr-qc]]}

\bibitem{bonnor}
   W. B. Bonnor,  {\it ``Static magnetic fields in general relativity''}, 
   \href{https://doi.org/10.1088/0370-1298/67/3/305}{Proc. Phys. Soc. Lond. A, 67:225–232, (1954)}

\bibitem{melvin}
   M. A. Melvin,  {\it ``Pure magnetic and electric geons''},  
   \href{https://doi.org/10.1016/0031-9163(64)90801-7}{Phys. Lett., 8:65–68, (1964)}

\bibitem{carminati}
N.~Van den Bergh and J.~Carminati,
{\it ``Non-aligned Einstein{\textendash}Maxwell Robinson{\textendash}Trautman fields of Petrov type D''},
\href{https://doi.org/10.1088/1361-6382/abbba3}{Class. Quant. Grav. \textbf{37} (2020) no.21, 215010};
\href{https://arxiv.org/pdf/2009.11516}{\tt [arXiv:2009.11516 [gr-qc]]} 


\bibitem{levi-civita}
T. Levi-Civita,
{\it ``Realt\`a fisica di alcuni spazi normali del Bianchi''}, Rend. Acc. Lincei, 26:519–531 (1917). \ {\it ``Republication of: The physical reality of some normal spaces of Bianchi''};
\href{https://doi.org/10.1007/s10714-011-1188-4}{Gen Relativ Gravit 43, 2307–2320 (2011)}.

\bibitem{bertotti}
B.~Bertotti,
{\it ``Uniform electromagnetic field in the theory of general relativity''}, \href{https://doi.org/10.1103/PhysRev.116.1331}{Phys. Rev. \textbf{116} (1959), 1331}

\bibitem{robinson}
I.~Robinson,
{\it ``A Solution of the Maxwell-Einstein Equations''},
Bull. Acad. Pol. Sci. Ser. Sci. Math. Astron. Phys. \textbf{7} (1959), 351-352

\bibitem{maccallum}
M.~M.~Akbar and M.~A.~H.~MacCallum,
{\it ``Static Axisymmetric Einstein Equations in Vacuum: Symmetry, New Solutions and Ricci Solitons''}
\href{https://doi.org/10.1103/PhysRevD.92.063017}{Phys. Rev. D \textbf{92} (2015) no.6, 063017};
\href{https://arxiv.org/pdf/1508.05196}{\tt [arXiv:1508.05196 [gr-qc]]}

\bibitem{PD-NUTs}
  M.~Astorino and G. Boldi,
     {\it ``Plebanski-Demianski goes NUTs (to remove the Misner string)''},
      \href{https://doi.org/10.1007/JHEP08(2023)085}{JHEP \textbf{08} (2023), 085};
       \href{https://arxiv.org/pdf/2305.03744.pdf}{\tt [arXiv:2305.03744 [gr-qc]]}

\bibitem{most-D}
M.~Astorino,
{\it ``Most general type-D black hole and the accelerating Reissner-Nordstrom-NUT-(A)dS solution''},
\href{https://doi.org/10.1103/PhysRevD.110.104054}{Phys. Rev. D \textbf{110} (2024) no.10, 104054};
\href{https://arxiv.org/pdf/2404.06551}{\tt [arXiv:2404.06551 [gr-qc]]}

\bibitem{revisiting-D}
H.~Ovcharenko, J.~Podolsky and M.~Astorino,
{\it ``Revisiting black holes of algebraic type D with a cosmological constant''},
\href{https://doi.org/10.1103/PhysRevD.111.084016}{Phys. Rev. D \textbf{111} (2025) no.8, 084016};
\href{https://arxiv.org/pdf/2501.07537}{\tt [arXiv:2501.07537 [gr-qc]]}


\bibitem{ernst2}
 F.~J.~Ernst,
  {\it ``New Formulation of the Axially Symmetric Gravitational Field Problem. II''},
    \href{https://doi.org/10.1103/PhysRev.168.1415}{\tt Phys.\ Rev.\  {\bf 168} (1968) 1415.}

\bibitem{swirling}
 M.~Astorino, R.~Martelli and A.~Vigan\`o,
  {\it ``Black holes in a swirling universe''},
   \href{https://doi.org/10.1103/PhysRevD.106.064014}{Phys. Rev. D \textbf{106} (2022) no.6, 064014};
    \href{https://arxiv.org/pdf/2205.13548}{\tt [arXiv:2205.13548 [gr-qc]]}.

\bibitem{harrison}
   B. Kent Harrison, {\it ``New Solutions of the Einstein‐Maxwell Equations from Old''}, 
\href{https://doi.org/10.1063/1.1664508}{J. Math. Phys. 9 (11): 1744–1752, (1968)}

\bibitem{ortaggio}
M.~Ortaggio and M.~Astorino,
{\it ``Ultrarelativistic boost of a black hole in the magnetic universe of Levi-Civita{\textendash}Bertotti{\textendash}Robinson''},
\href{https://doi.org/10.1103/PhysRevD.97.104052}{Phys. Rev. D \textbf{97} (2018) no.10, 104052};
\href{https://arxiv.org/pdf/1805.05382}{\tt [arXiv:1805.05382 [gr-qc]]}

\bibitem{charging}
M.~Astorino,
{\it ``Charging axisymmetric space-times with cosmological constant''},
\href{https://doi.org/10.1007/JHEP06(2012)086}{JHEP \textbf{06} (2012), 086} ;
\href{https://arxiv.org/pdf/1205.6998.pdf}{\tt [arXiv:1205.6998 [gr-qc]]}

\bibitem{illy}
M.~Illy,
{\it ``Accelerated Reissner-Nordstrom black hole in a swirling, magnetic universe''}, \href{https://arxiv.org/pdf/2312.14995}{\tt [arXiv:2312.14995 [gr-qc]]}.

\bibitem{dipinto}
A.~Di Pinto, S.~Klemm and A.~Vigan{\`o},
{\it ``Kerr-Newman black hole in a Melvin-swirling universe''},
\href{https://doi.org/10.1007/JHEP06(2025)150}{JHEP \textbf{06} (2025), 150}, 
\href{https://arxiv.org/pdf/2503.07780}{\tt[arXiv:2503.07780 [gr-qc]]}


\bibitem{static-typeI-bh}
M.~Astorino,
{\it ``Static hairy black hole in 4D General Relativity''}, (2025); 
\href{https://doi.org/10.6084/m9.figshare.30081115.v1}{\tt https://doi.org/10.6084/m9.figshare.30081115.v1}

\bibitem{Type-I}
M.~Astorino,
 {\it ``Accelerating and Charged Type I Black Holes''},
  \href{https://doi.org/10.1103/PhysRevD.108.124025}{Phys. Rev. D \textbf{108} (2023) no.12, 124025};
   \href{https://arxiv.org/pdf/2307.10534.pdf}{\tt [arXiv:2307.10534 [gr-qc]]}

\bibitem{stephani-big-book}
  H.~Stephani, D.~Kramer, M.~A.~H.~MacCallum, C.~Hoenselaers and E.~Herlt,
  {``Exact solutions of Einstein's field equations''}, \ 
  \href{https://doi.org/10.1017/CBO9780511535185}{\tt [doi:10.1017/CBO9780511535185]}

\bibitem{enhanced}
  M.~Astorino,
  {\it ``Enhanced Ehlers Transformation and the Majumdar-Papapetrou-NUT Spacetime''},
  \href{https://doi.org/10.1007/JHEP01(2020)123}{JHEP \textbf{01} (2020), 123}; 
   \href{https://arxiv.org/pdf/1906.08228}{\tt [arXiv:1906.08228 [gr-qc]]}

\bibitem{kerr-melvin-mass}
M.~Astorino, G.~Comp\`ere, R.~Oliveri and N.~Vandevoorde,
{\it ``Mass of Kerr-Newman black holes in an external magnetic field''},
\href{https://doi.org/10.1103/PhysRevD.94.024019}{Phys. Rev. D \textbf{94} (2016) no.2, 024019}; 
\href{https://arxiv.org/pdf/1602.08110}{\tt [arXiv:1602.08110 [gr-qc]]}.

\bibitem{Podolsky:2018dpr}
J.~Podolsky, O.~Hru{\v{s}}ka and J.~B.~Griffiths,
{\it ``Non-expanding Pleba{\'n}ski{\textendash}Demia{\'n}ski space-times''},
\href{https://doi.org/10.1088/1361-6382/aacdd5}{Class. Quant. Grav. \textbf{35} (2018) no.16, 165011}; 
\href{https://arxiv.org/pdf/1804.01519}{\tt [arXiv:1804.01519 [gr-qc]]}

\bibitem{witten}
E.~Witten,
{\it ``Instability of the Kaluza-Klein Vacuum''},
\href{https://doi.org/10.1016/0550-3213(82)90007-4}{Nucl. Phys. B \textbf{195} (1982), 481-492}

\bibitem{bh+bubble}
 M.~Astorino, R.~Emparan and A.~Vigan\`o,
 {\it ``Bubbles of nothing in binary black holes and black rings, and viceversa''},
  \href{https://doi.org/10.1007/JHEP07(2022)007}{JHEP \textbf{07} (2022), 007} ;
   \href{https://arxiv.org/pdf/2204.09690.pdf}{\tt [arXiv:2204.09690 [hep-th]]}.


\bibitem{thermo-acc}
M.~Astorino,
{\it ``Thermodynamics of Regular Accelerating Black Holes''},
\href{https://doi.org/10.1103/PhysRevD.95.064007}{Phys. Rev. D \textbf{95} (2017) no.6, 064007}; 
\href{https://arxiv.org/pdf/1612.04387.pdf}{\tt [arXiv:1612.04387 [gr-qc]]}

\bibitem{ruffini}
D.~Christodoulou and R.~Ruffini,
{\it ``Reversible transformations of a charged black hole''},
\href{https://doi.org/10.1103/PhysRevD.4.3552}{Phys. Rev. D \textbf{4} (1971), 3552-3555}

\bibitem{melvin-bertotti}
M.~Halilsoy and S.~H.~Mazharimousavi,
{\it``Unified Bertotti-Robinson and Melvin Spacetimes''},
\href{https://doi.org/10.1103/PhysRevD.88.064021}{Phys. Rev. D \textbf{88} (2013), 064021}; 
\href{https://arxiv.org/pdf/1211.6983}{\tt[arXiv:1211.6983 [gr-qc]]}.

\bibitem{kerr-newman-bertotti}
H.~Ovcharenko and J.~Podolsk{\'y},
{\it ``A new class of rotating charged black holes with non-aligned electromagnetic field''}, 
\href{https://arxiv.org/pdf/2508.04850}{\tt [arXiv:2508.04850 [gr-qc]]}

\bibitem{podolsky-book}
J.~B.~Griffiths and J.~Podolsky,
{\it ``Exact Space-Times in Einstein's General Relativity''}, 
\href{https://doi.org/10.1017/CBO9780511635397}{Cambridge University Press, 2009}

\bibitem{marcoa-embedding}
M.~Astorino,
{\it ``Embedding hairy black holes in a magnetic universe''},
\href{https://doi.org/10.1103/PhysRevD.87.084029}{Phys. Rev. D \textbf{87} (2013) no.8, 084029} ;
\href{https://arxiv.org/pdf/1301.6794}{\tt [arXiv:1301.6794 [gr-qc]]}.

\bibitem{herdeiro}
A.~Bokuli{\'c} and C.~A.~R.~Herdeiro,
{\it ``Generalised Harrison transformations and black diholes in Einstein-ModMax''}, 
\href{https://arxiv.org/pdf/2507.16926}{\tt [arXiv:2507.16926 [gr-qc]]}

\end{thebibliography}
\end{document}